\documentclass[12pt]{iopart}

\usepackage{bbold,amsmath,amssymb,latexsym,multirow,epsfig,stmaryrd}

    \PassOptionsToPackage{numbers, compress}{natbib}
\usepackage{natbib}
\bibpunct[, ]{(}{)}{,}{a}{}{,}%
\usepackage{iopams}  
\usepackage{graphicx}

\usepackage[utf8]{inputenc} % allow utf-8 input
\usepackage[T1]{fontenc}    % use 8-bit T1 fonts
\usepackage{hyperref}      % hyperlinks
\hypersetup{
    colorlinks=true,
    linkcolor=blue,
    filecolor=magenta,      
    urlcolor=cyan,
    pdftitle={Overleaf Example},
    pdfpagemode=FullScreen,
    }
\usepackage{url}            % simple URL typesetting
\usepackage{booktabs}       % professional-quality tables
\usepackage{amsfonts}       % blackboard math symbols
\usepackage{nicefrac}       % compact symbols for 1/2, etc.
\usepackage{microtype}      % microtypography
\usepackage{xcolor}         % colors
\usepackage[capitalize]{cleveref}
\usepackage{wrapfig}

\setcitestyle{numbers,open={(},close={)}} %Citation-related commands

\usepackage[english]{babel}

\def \< {\langle}
\def \> {\rangle}

\def \P {\mathbb{P}}

\def \S {\Sigma}

\def\be{\begin{equation}}
\def\ee{\end{equation}}
\def\bes{\begin{subequations}}
\def\ees{\end{subequations}}
\def\bea{\begin{eqnarray}}
\def\eea{\end{eqnarray}}
\def\bry{\begin{array}}
\def\ery{\end{array}}
\def\bit{\begin{itemize}}
\def\eit{\end{itemize}}
\def\ben{\begin{enumerate}}
\def\een{\end{enumerate}}

\def\({\left(}
\def\){\right)}

\def\S{\mathcal{S}}

\usepackage{fancyhdr}
\begin{document}

\title{The Universal Statistical Structure and Scaling Laws of Chaos and Turbulence}

\author{Noam Levi~~and Yaron Oz}
\address{Raymond and Beverly Sackler School of Physics and Astronomy, 
  Tel-Aviv University, Tel-Aviv 69978, Israel}
\ead{noam@mail.tau.ac.il}

\begin{abstract}
Turbulence is a complex spatial and temporal structure created by the strong non-linear dynamics of fluid flows at high Reynolds numbers. Despite being an ubiquitous phenomenon that has been studied for centuries, a full understanding of turbulence remained a formidable challenge.  Here, we introduce tools from the fields of quantum chaos and Random Matrix Theory (RMT) and present a detailed analysis of image datasets generated from turbulence simulations of incompressible and compressible fluid flows.
Focusing on two observables: the data Gram matrix and the single image distribution, we study both the local and global eigenvalue statistics and
compare them to classical chaos, uncorrelated noise and natural images. 
We show that from the RMT perspective, the turbulence Gram matrices lie in the same universality class as quantum chaotic rather than integrable systems, and the data exhibits power-law scalings in the bulk of its eigenvalues which are vastly different from uncorrelated classical chaos, random data, natural images. Interestingly, we find that the single sample distribution only appears as fully RMT chaotic, but deviates from chaos at larger correlation lengths, as well as exhibiting different scaling properties.

\end{abstract}

\noindent\hrulefill
\tableofcontents
\noindent\hrulefill

\section{Introduction}

Understanding the space of solutions to the incompressible Navier-Stokes (NS) equations, is one of the Millennium Prize Problems in mathematics. The strong dynamics in the highly non-linear regime generates a complex spatial and temporal structure of fluid turbulence \cite{Frisch},
and single realization of a turbulent solution to the NS equations is unpredictable.  Studying statistical averages of fluid velocity observables seems to reveal a hidden
scaling structure at the inertial range \cite{Kolmogorov}, yet learning 
this distribution remained a challenge, which is widely considered as the most important unsolved problem in classical physics. A complete understanding of turbulence statistics is expected to provide valuable insights to the dynamics of diverse strongly interacting physical systems that are far from thermal equilibrium.

The incompressible NS equations provide a mathematical formulation of the fluid flow evolution
at velocities much smaller than the speed of sound: 

\begin{equation}
\partial_t v^i + v^j\partial_j v^i =
-\partial^i p + \nu \partial_{jj} v^i  + f^i,~~~~~~\partial_iv^i = 0  \ ,
\label{NS}
\end{equation}
where $v^i, i = 1,\ldots,d$  is the fluid velocity in $d$ space dimensions, $p$ is the fluid pressure, $\nu$ is the kinematic viscosity and $f^i$ is an external
random force. 
An important dimensionless parameter in the study of fluid flows  is the Reynolds number 
${\cal R}_e = \frac{l v}{\nu}$, where $l$
is a characteristic length scale, $v$ is the velocity difference at that scale, and $\nu$ is the kinematic viscosity. 
The Reynolds number quantifies the relative strength of the
non-linear interaction compared to the viscous term in (\ref{NS}).
When the Reynolds number is of order $10-10^2$ one observes a chaotic 
fluid flow, while when it is $10^3$ or higher,
one observes a fully developed turbulent structure of the flow.

\begin{figure}
    \centering
    \includegraphics[width=1\textwidth]{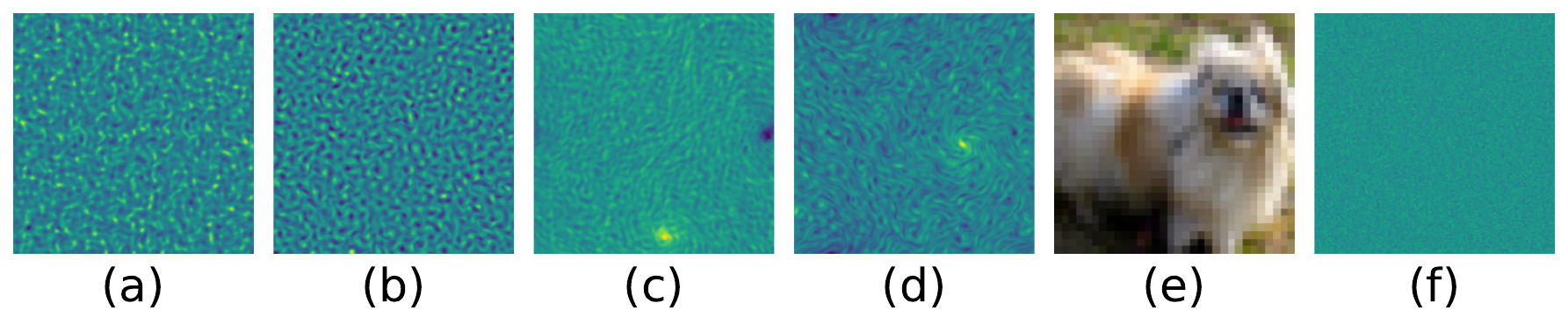}
    \caption{
    {\bf Left to Right:} Images of vorticity:
    compressible chaos, incompressible chaos, compressible turbulence, incompressible turbulence; CIFAR-10 and uncorrelated Gaussian data. The turbulence and chaos data are solutions of the NS equations  from \cite{whittaker2023neural}.
      }
    \label{fig:images}
\end{figure}
We will consider the case of two spatial dimensions
\cite{doi:10.1063/1.1762301,boffetta2012two} and use the vorticity pseudoscalar $\omega=\epsilon_{ij} \partial^iv^j$ to recast (\ref{NS}) as
\begin{align}
    \label{eq::vort}
    \partial_t \omega =   - v^i \partial_i\omega + \nu \partial_{jj} \omega + \epsilon_{ij}\partial^if^j \ .
\end{align}
Using divergence-free and statistically homogeneous and isotropic
Gaussian random forcing function, one
generates a dataset of two-dimensional incompressible fluid flows (Fig.1d) by numerically evolving the vorticity equation (\ref{eq::vort}) \cite{whittaker2023neural}.  
The chaotic dataset (Fig.1b) is defined 
by the set of snapshots in the time steps before turbulence scaling is observed. 
Similarly, one can simulate weakly compressible fluid flows (Fig.1a,c) \cite{whittaker2023neural}, where 
the fluid density $\rho$ is not constant and the velocity vector field is not solenoidal. 
The simulations provide high dimensional images, composed of $435\times435$ pixels.

Deep learning has emerged as a powerful tool with the potential to address  complex statistical problems, hence
harnessing its capabilities for the analysis of turbulence seems imperative. 
One such avenue is the utilization of deep learning methods to generate turbulent flows or single-particle trajectories in turbulence \cite{Drygala_2022,tretiak2022physics,Shu_2023, yang2023denoising,li2023synthetic,whit}, 
 bypassing the need for numerical simulations and experiments to improve the precision of statistical turbulence. 
In this work we initiate a new route in the study of turbulence
by employing tools from the quantum chaos and Random Matrix Theory (RMT) to study and characterize both the local and global eigenvalue structure 
of the data Gram matrix and the single image distribution of turbulence in comparison to noise, real world images (Cifar-10~\cite{cifar10}, $32\times 32$ single channel) and classical chaos.
We show that turbulence data covariance and single image matrices exhibit power-law eigenspectrum scalings that differ from uncorrelated classical chaos, random data, natural images.
We further find that the turbulence Gram matrices lie in the universality class of quantum chaotic systems,
while the single sample distribution deviates from this class at larger correlation lengths, thus indicating different levels of ergodicity.
We will discuss some of the implications of these results in the conclusions section.

\section{Background and Related Work}
\label{sec:related_app}

% Here, we provide some additional background material, complimenting the analysis performed in the main text. 

\paragraph{Neural Scaling Laws}
The so-called neural scaling laws constitute a set of empirical findings elucidating the interrelations between neural network size, training data, computing resources, and performance. Initiated in \cite{kaplan2020scaling}, these lawful connections have since been substantiated through investigations by \cite{maloney2022solvable, hernandez2022scaling} among others, and further scrutinized in \cite{DBLP:conf/emnlp/IvgiCB22, DBLP:conf/nips/AlabdulmohsinNZ22, DBLP:journals/jmlr/SharmaK22, DBLP:conf/nips/SorscherGSGM22, DBLP:journals/corr/abs-2302-09049, DBLP:journals/corr/abs-2302-09650}. The crux is that test error decreases as a power law of parameters in a predictable manner, underlied by the behavior of the data on which the networks were trained. However, this relationship eventually attenuates. The conundrum is that myriad interacting factors underlie network behavior, obscuring the mechanisms behind the scaling. Insights have been gained by applying random matrix theory.

\paragraph{Random Matrix Theory}
Random matrix theory originated in studies of large random matrices. It is best suited to analyzing numerous realizations of high-dimensional systems with copious samples, such that their dimension-to-samples ratio is finite. Although first applied to random matrices, it has proven fruitful more broadly in machine learning, e.g. for nonlinear regression~\cite{pennington2017nonlinear}, Fourier models~\cite{Liao_2021}, Hessian spectra~\cite{liao2021hessian}, and weight statistics~\cite{martin2019traditional,Thamm}. For a survey see \cite{couillet_liao_2022}.

\paragraph{Universality}
Much effort has focused on universality – the emergence of common features in diverse systems when sufficiently large. For instance, eigenvalue spectra from disparate random processes can exhibit similar distributions. Universality is potent because a simple tractable System B can illuminate a complex intractable System A if they belong to the same universality class~\cite{Bao_2015,baik2004phase,hu2022universality,bai2010spectral}. In our work, we refer to the notion that System A represents real-world datasets with opaque statistics, while System B is a random matrix with a special correlation structure, studied in~\cite{levi2023underlying}. The observation that the data from dynamical systems aligns with RMT predictions allows us to leverage the simplicity of RMT to gain insights into the complex statistics of the former.
Our methods for demonstrating universal properties are taken from the Quantum Chaos literature, expounded upon in~\cite{Kim_2023,pandey1983random,Liu_2018,vidmar_1905.06345}.

\section{Statistical Structure of Natural Datasets}
\label{sec:datasets}

We consider two classes of observables constructed from the same samples for each dataset. 
The first class is obtained by defining $X \in \mathbb{R}^{d\times M}$, where $d$ is the dimension of each flattened image vector and $M$ is the number of samples. We then compute the empirical feature-feature covariance (Gram) matrix, $\Sigma_M = \tfrac{1}{M} X X^T$. This observable represents an average over the data, holds no spatial information, and is expected to follow certain Central Limit Theorem behaviors.
The second class is defined by $\chi\in \mathbb{R}^{d_x \times d_y}$, where $d_x,d_y$ are the physical dimensions of the images in the data and $d_x\times d_y=d$. Here, $\chi$ represents a single sample, from which we compute the product $\Sigma_\chi= \chi \chi^T \in \mathbb{R}^{d_x \times d_x}$, whose eigenvalues are the squared singular values of $\chi$. By generating $\Sigma_\chi$ for each sample, and studying the distribution of them all, we will access the single image distribution, without any averaging. In our work, $d_x=d_y=435$ for all datasets apart from Cifar-10.

% \subsection{Random Matrix Theory and Quantum Chaos}

\subsection{Particular Global Structure}
\label{sec:global}

The properties of $\Sigma_M$ and $\Sigma_\chi$ in natural image data are entirely unknown a priori, as we do not know how to parameterize the process which generated natural images. Nevertheless, interesting observations have been made. Empirical evidence shows that the spectrum of $\Sigma_M$ for various datasets can be separated into a set of large eigenvalues ($\mathcal{O}(10)$), a bulk of eigenvalues which decay as a power law $\lambda_i \sim i^{-1-\alpha}$ and a large tail of small values which terminates at some finite index $n$. Since the top eigenvalues represent the largest overlapping properties across different samples, these are not simply interpreted without more information on the underlying distribution. The bulk of the eigenvalues, however, can be understood as representing the correlation structure of different features amongst themselves, and has been key to understanding the emergence of neural scaling laws~\cite{kaplan2020scaling, maloney2022solvable}. 

\begin{figure}[t!]
    \centering
    \includegraphics[width=1\textwidth]{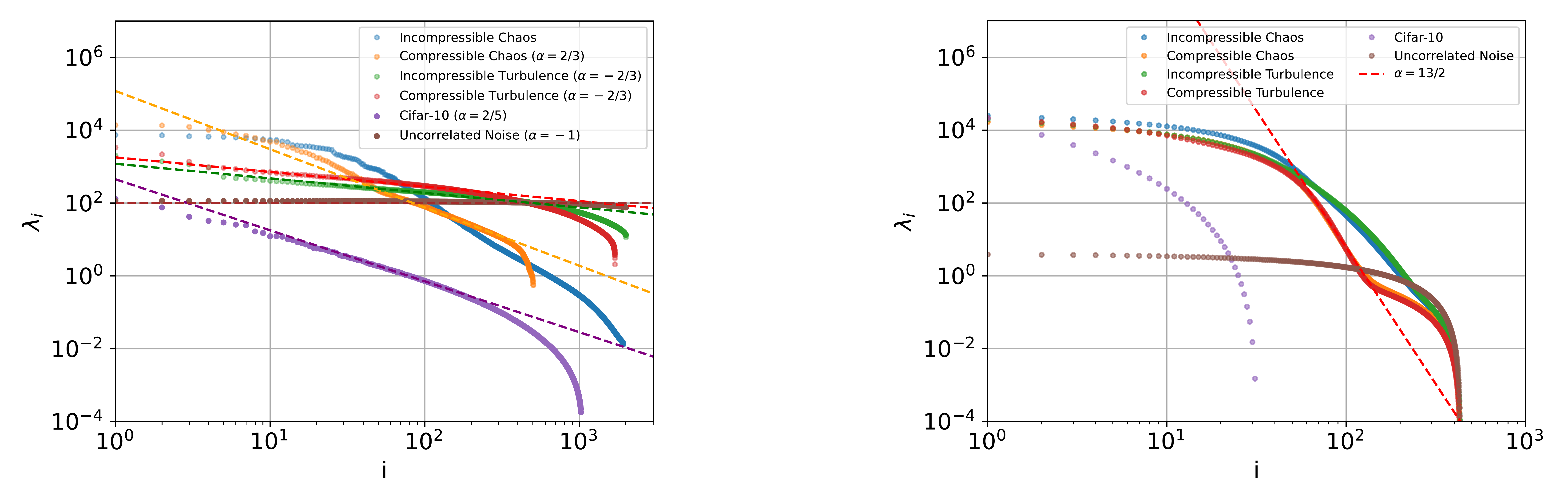}
    \caption{ 
 {\bf Left:}  Scree plot of the Gram matrix $\Sigma_{M}$ for several datasets, as well as for uncorrelated noise. We observe distinct scaling laws for the eigenvalue bulk of different datasets. Natural images, as well as turbulence, scale as $\lambda_i \propto i^{-1-\alpha}$, while chaos appears to be multi scaled, composed of several decaying bulks with different $\alpha$ values. We note that noise does not scale, as expected.
 {\bf Right:} 
 Scree plot of the single sample matrix $\Sigma_{\chi}$ for the same datasets. We observe a significantly different behavior for the eigenvalue bulk. We see that at the single sample level, the separation between compressible and incompressible fluids manifests as much more important than the turbulence/chaos distinction. 
 In both figures the number of samples is taken to be $M=2000$.
 }
    \label{fig:gram_vs_Single_Sample_scaling}
\end{figure}
\vspace{2.cm}
In \cite{levi2023underlying}, the bulk behavior was interpreted by appealing to a simple model of correlated Gaussian data, where the population covariance $\Sigma$ was taken to be
\begin{equation}
\label{eq:toeplitz}
    \Sigma_{ij}^\mathrm{Toe}
    =
    S, \qquad
    T
    = I_{ij} + c |i-j|^\alpha = U^\dagger S V,
    \qquad \alpha,c \in \mathbb{R}.
\end{equation}
The matrix $\Sigma_{ij}^\mathrm{Toe}$ is a diagonal matrix of singular values $S$ constructed from $T$, a full-band Toeplitz matrix. The sign of $\alpha$ dictates whether correlations decay (negative) or intensify (positive) with distance along a one-dimensional feature space.

\begin{figure}
    \centering
    \includegraphics[width=.32\textwidth]{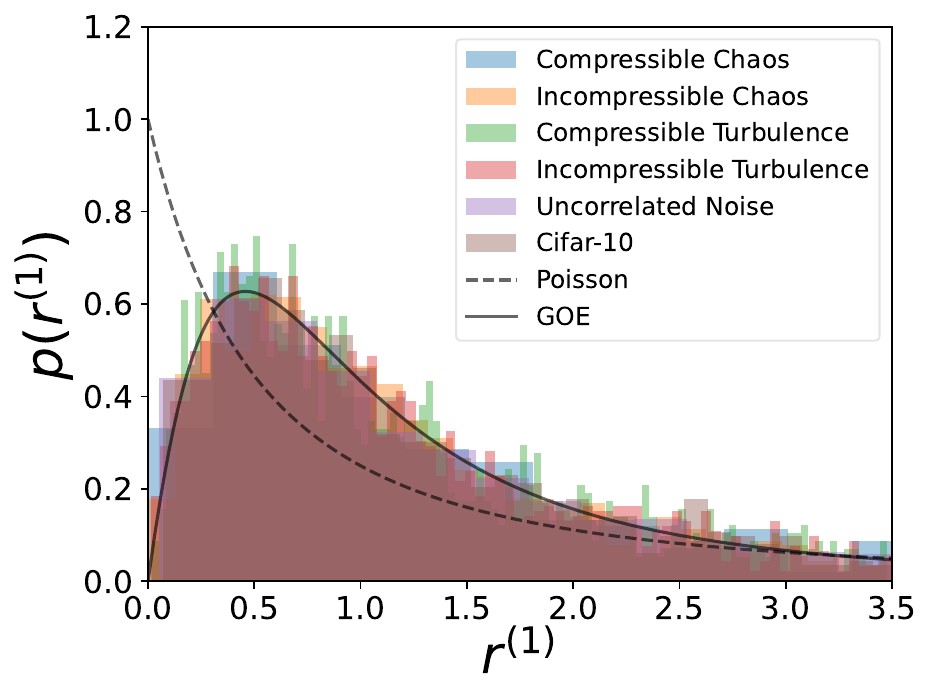}
    \includegraphics[width=.32\textwidth]{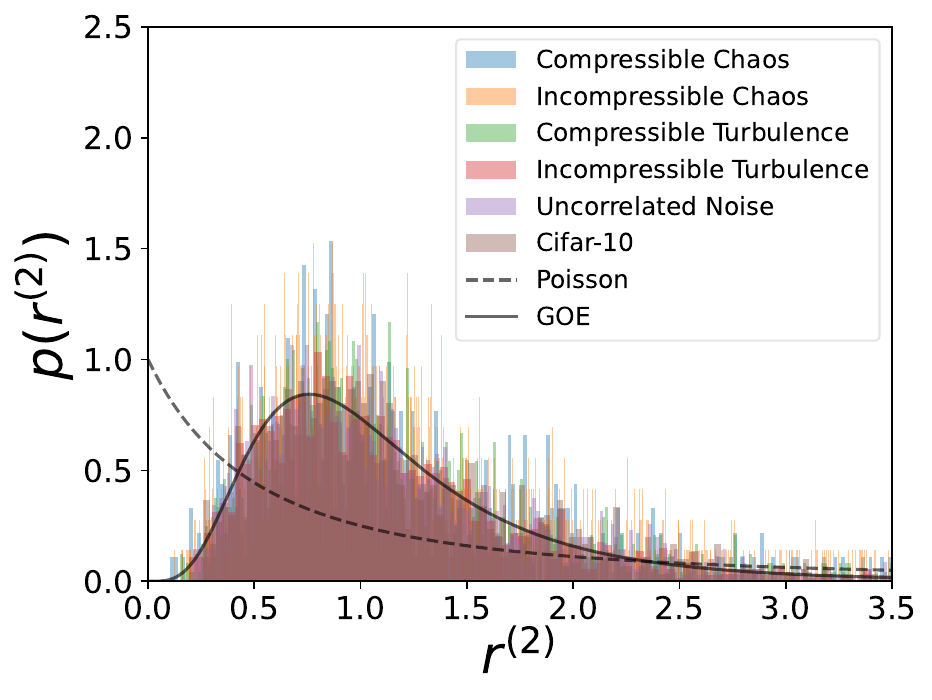}
    \includegraphics[width=.32\textwidth]{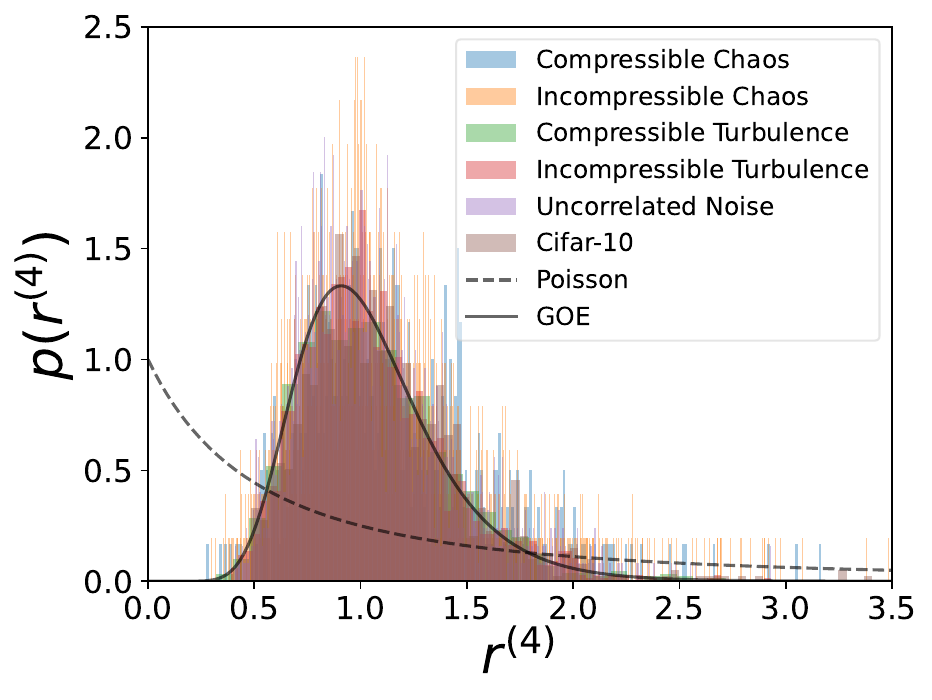}
    \includegraphics[width=.32\textwidth]{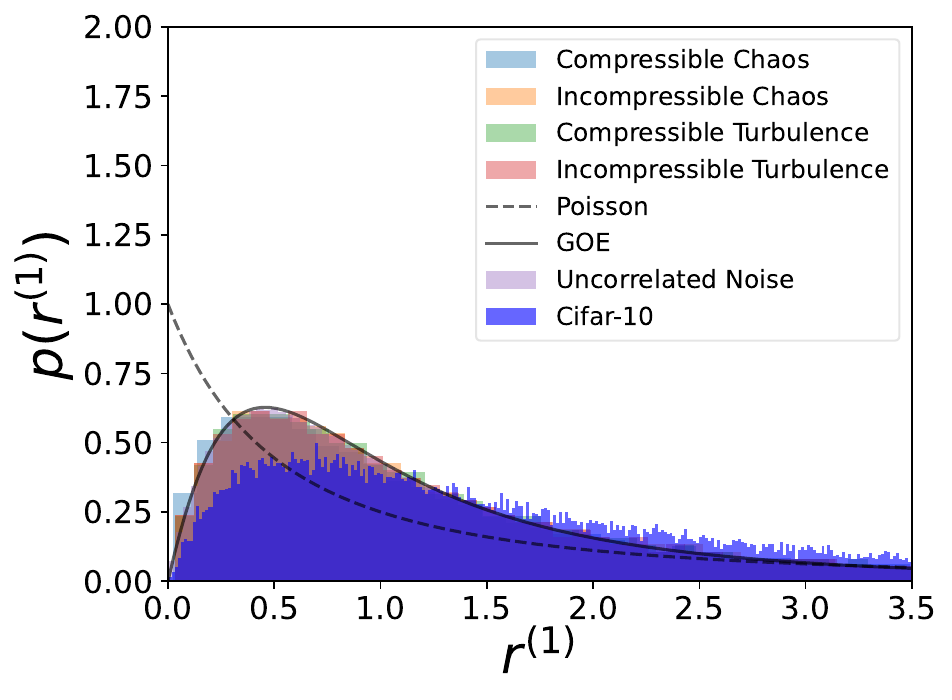}
    \includegraphics[width=.32\textwidth]{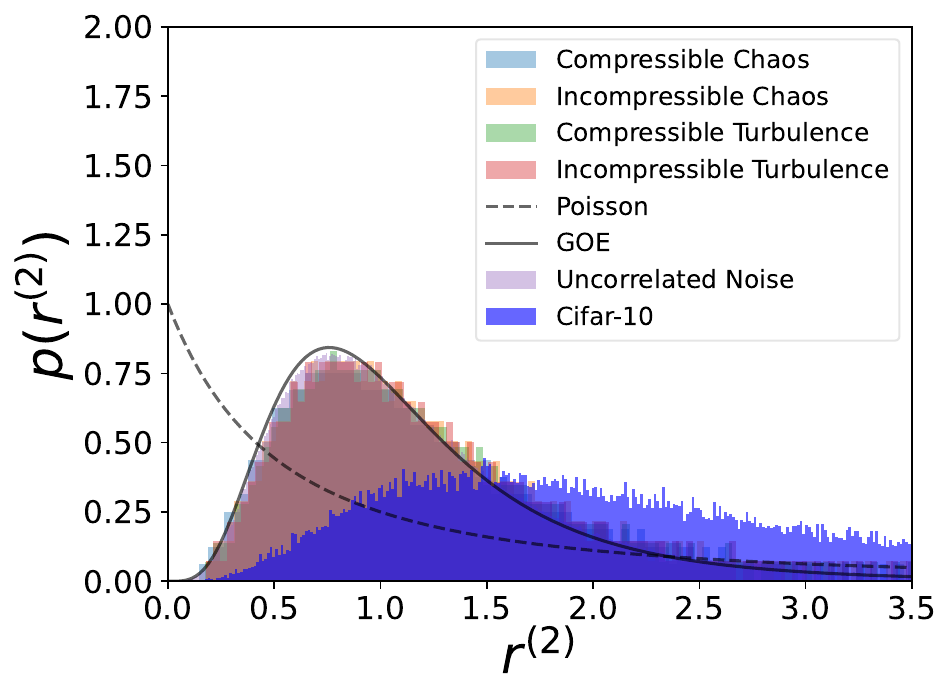}
    \includegraphics[width=.32\textwidth]{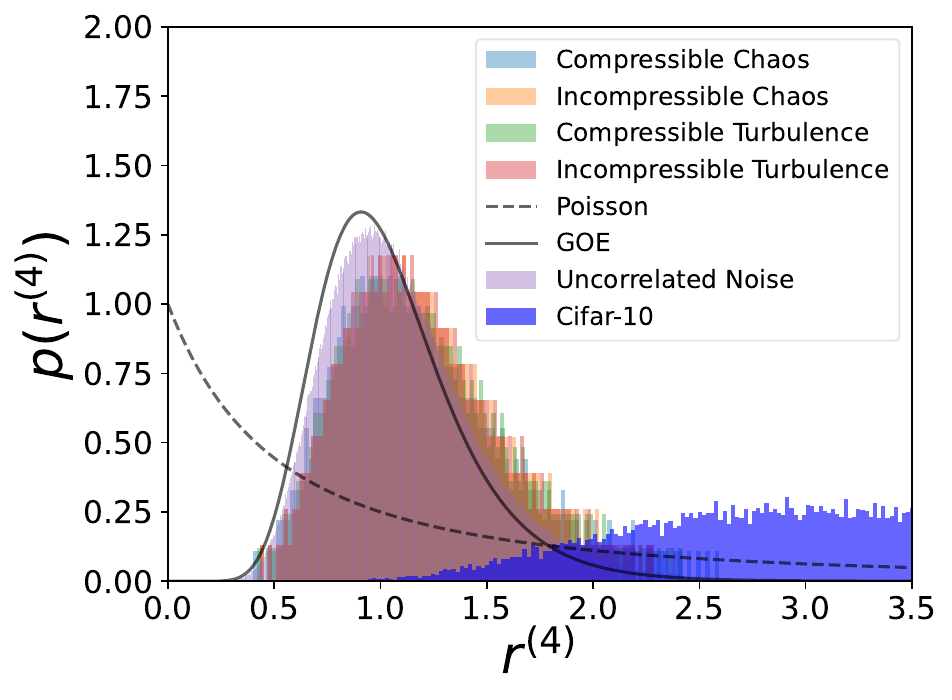}    
\caption{
     {\bf Left to Right:} The $r^{(n)}$ probability densities for $n=1,2,4$, for the fluid equations' solutions as well as natural data (Cifar-10) and uncorrelated noise.
    Black curves indicate the RMT predictions for the GOE distribution from~\cref{eq:r_dist}. 
    {\bf Top row:} The results for the Gram matrices $\Sigma_M$. 
    {\bf Bottom row:} The same distributions for the single sample matrices $\Sigma_\chi$.
    The top panels indicate that the bulk of any Gram matrix eigenvalues under investigation belongs to the GOE universality class, and that system has enough statistics to converge to the RMT predictions.
    Conversely, the bottom row demonstrates that in terms of the $r^{(1)}, r^{(2)}$ distribution, the dynamical systems appear to be GOE, as does uncorrelated noise, while natural images are clearly different. On larger separation scales, captured by $r^{(4)}$, we see that the fluid solutions begin to diverge from GOE behavior, indicating that the system is not fully ergodic at the single sample level. 
    }
        % \vspace{-.5cm}
    \label{fig:Single_Sample_scaling}
\end{figure}

In \cref{fig:gram_vs_Single_Sample_scaling}, we show the eigenvalue scaling of both the Gram matrices (left) and the single images distribution (right), for compressible and incompressible turbulence/chaos, Cifar-10 images, and uncorrelated noise.
We observe several interesting properties. Firstly, the Gram matrices of natural datasets and turbulence display a single power law scaling for its bulk, while the chaotic data demonstrates multiple exponents, related to correlations at different scales. This trait makes the two datasets distinguishable at the second moment level. Secondly, natural data typically has $0<\alpha<1/2$~\cite{levi2023underlying}, implying positive correlations which increase in distance, while the turbulent fluid has negative $\alpha\simeq -2/3$, in between natural images and uncorrelated noise ($\alpha=-1$), which implies decaying correlations between features. Lastly, we find a clear difference between the single image statistics and the Gram matrix for turbulence and chaos. In the single image case, both chaos and turbulence seem to share nearly the same eigenvalue scaling structure, and the difference comes from the compressible and incompressible property of the fluid itself.

\subsection{Universal Local Structure}
\label{sec:local}

\begin{figure}[ht!]
    \centering
    \includegraphics[width=0.5\textwidth]{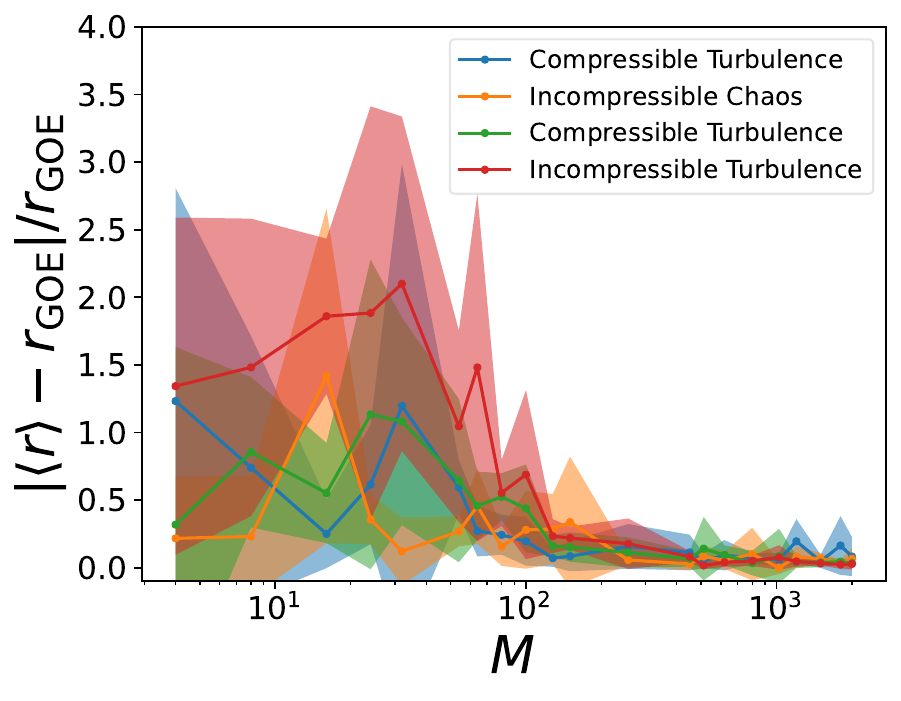}
    \caption{ 
    Convergence of the NS solutions' Gram matrices $\Sigma_M$ to the RMT regime. We compare the average $\langle r \rangle$ to the GOE value $\langle r \rangle_\mathrm{GOE}\simeq 7/4$. 
    Convergence occurs at $M\sim 200$, while the sample dimensions are $435\times435$, implying that convergence to RMT is very fast. {\it Solid} curves are computed by averaging over 5 sampling iterations, and the {\it shaded} regions represent one standard deviation.
    }
    \vspace{-.5cm}
    \label{fig:r_converge}
\end{figure}

The $r$-statistics, first introduced in \cite{huse_r_stat}, a diagnostic tool for short-range correlations, defined directly on the spectrum.
We define ratios of spacings between eigenvalues $\cdots < \lambda_i < \lambda_{i+1} < \cdots$ as
\begin{align}
\label{eq:ri_definition}
& r_i^{(n)} =
\frac{\lambda_{i+2n} - \lambda_{i+n}  } 
{\lambda_{i+n} - \lambda_{i} }
\ .
\end{align}
The $r^{(n)}_i$ distribution takes unique values if the spectra are the eigenvalues of random matrices:
\begin{align}
\label{eq:r_dist}
p_\mathrm{GOE}(r^{(n)}=r)
=
Z_\nu
\frac{(r+r^2)^\nu}{(1+r+r^2)^{1+3/2\nu} },
\quad
p_\mathrm{Pois}(r^{(n)}=r)
=
\frac{r^{n-1}}{(1+r^2)^{2n}}, 
\end{align}
where $\beta=1$, $Z_\nu$ is a normalization constant, and $\nu = \frac{n(n+1)}{2} \beta + n-1$. The expectation value of the ratios $r_i=r^{(1)}_i$ for matrices in the GOE is $\langle r\rangle \approx 7/4=1.75$, while it diverges for integrable systems, approaching $\langle r \rangle \to \infty$ for a Poisson process \cite{r_ratios_theory_values}. In~\cref{fig:Single_Sample_scaling} (top row), we show the $r^{(n)}, n=1,2,4$ distributions for the Gram matrices $\Sigma_M$ of dynamical systems, against Cifar-10 data and uncorrelated noise. Clearly, all datasets converge to the GOE predictions, demonstrating that Gram matrices are universally GOE. In \cref{fig:r_converge} we show this convergence occurs at very low relative sample number.
The bottom row of \cref{fig:Single_Sample_scaling} shows the same distributions for the single sample $\Sigma_\chi$. Here, we see a clear distinction between natural images (Cifar-10) and the dynamical simulations. Clearly, the dynamical system exhibits a higher level of ergodicity
than Cifar-10.

\section{Conclusions}

Statistical turbulence exhibits scaling exponents of fluid observables at the inertial range of scales
$l \ll r \ll L$, where the distance scales $l$ and $L$ are determined in terms of the viscosity and driving force, respectively.  
Denote the velocity of the fluid by $\vec{v}(t,\vec{r})$, then the turbulent behavior can be characterized by the longitudinal structure functions $S_n(r) = \langle (\delta v(r))^n \rangle$ of velocity differences 
%$\delta v(r)  = (\vec{v}(\vec{r}) - \vec{v}(0))\cdot \frac{\vec{r}}{|\vec{r}|}$%
between points separated by a fixed distance $r$.
$S_n(r) \sim r^{\xi_n}$,
where the exponents $\xi_n$  are independent of the fluid
details and depend only on the number of spatial dimensions \cite{Frisch}. In particular, $\xi_2 \simeq \frac{2}{3}$ 
characterizes the fluid energy spectrum.
We observed a seemingly unrelated scaling of the bulk eigenvalues
of the Gram covariance matrix (curiously $\alpha \simeq -\xi_2$). It would be interesting to gain a better understanding of this new scaling from a dynamical viewpoint of turbulence. Further, it will be valuable to know whether the
eigenspectrum scaling depends on the number of space dimensions as is the case with $\xi_n$.
In general, $S_n(r)$ scaling is expected to be independent of whether it is calculated
from an ensemble average, 
or from one sample and averaging over points separated by a fixed distance $r$. Our results
show that the Gram matrix exhibits higher level of ergodicity, unlike the case of one sample. This can be used to quantify the 
effect of the finite size sample on the accuracy of turbulence scalings,
which the higher structure functions are expected to detect.
\label{sec:conclusions}

\section*{Acknowledgments}

We would like to thank Tim Wittaker for sharing with us 
his numerical solutions of the Navier-Stokes equations 
in \cite{whittaker2023neural}. This work is supported by the ISF Center of Excellence.
 N.L. would like to thank the Milner Foundation for the award of a Milner Fellowship.
% The work of Y.O. is supported in part by Israel Science Foundation Center of Excellence.
% This work was performed in part at Aspen Center for Physics, which is supported by the U.S. National Science Foundation grant PHY-2210452.

\bibliography{turb.bib}
% \bibliographystyle{unsrt}

% \appendix

\end{document}